\RequirePackage{fix-cm}
\documentclass[prl,aps,reprint,floatfix,superscriptaddress]{revtex4-2}
\usepackage{amsmath,amssymb,amsfonts}
\usepackage{slashed}
\usepackage{graphicx}
\usepackage{bm}
\usepackage[colorlinks,linkcolor=blue,citecolor=blue]{hyperref}

\newcommand{\be}{\begin{equation}}
\newcommand{\ee}{\end{equation}}
\newcommand{\bea}{\begin{eqnarray}}
\newcommand{\eea}{\end{eqnarray}}
\newcommand{\ba}{\begin{eqnarray}}
\newcommand{\ea}{\end{eqnarray}}

\def\be{\begin{eqnarray}}
\def\ee{\end{eqnarray}}
\def\bea{\be}
\def\eea{\ee}

\def\roughly#1{\mathrel{\raise.3ex\hbox{$#1$\kern-.75em%
\lower1ex\hbox{$\sim$}}}}

\renewcommand{\th}{\theta}
\newcommand{\dl}{\delta}

\newcommand{\eps}{\varepsilon}
\newcommand{\f}[2]{\frac{#1}{#2}}

\newcommand{\chid}{\chi^\dagger}

\newcommand{\eq}{\begin{equation}}
\newcommand{\eqx}{\end{equation}}
\newcommand{\eqn}{\begin{eqnarray}}
\newcommand{\eqnx}{\end{eqnarray}}

\begin{document}

\title{Universality and emergent effective fluid from jets and string breaking \\ in the massive Schwinger model using tensor networks}

\author{Romuald A. Janik}
\email{romuald.janik@gmail.com}

\author{Maciej A. Nowak}
\email{maciej.a.nowak@uj.edu.pl}

\author{Marek M. Rams}
\email{marek.rams@uj.edu.pl}
\affiliation{Institute of Theoretical Physics and Mark Kac Center for Complex Systems Research, Jagiellonian University, 30-348 Krak\'{o}w, Poland}

\author{Ismail Zahed}
\email[]{ismail.zahed@stonybrook.edu}
\affiliation{Center for Nuclear Theory, Department of Physics and Astronomy, Stony Brook University, Stony Brook, New York 11794–3800, USA}

\date{\today}
\begin{abstract}
We analyze the correlation between the energy, momentum and spatial entanglement produced by two luminal jets in the massive Schwinger model. Using tensor network methods, we show that for $m/g>1/\pi$, in the vicinity of the strong to weak coupling transition, a nearly perfect and chargeless effective fluid behavior appears around the mid-rapidity region with a universal energy-pressure relationship.
The evolution of energy and pressure is strongly correlated with the rise of the spatial entanglement entropy, indicating a key role of quantum dynamics.
Some of these observations may be used to analyze high multiplicity jet fragmentation events, energy-energy and energy-charge correlators at current collider energies.
\end{abstract}

\maketitle

\begin{figure}[t]
\begin{center}
\includegraphics[width=\columnwidth]{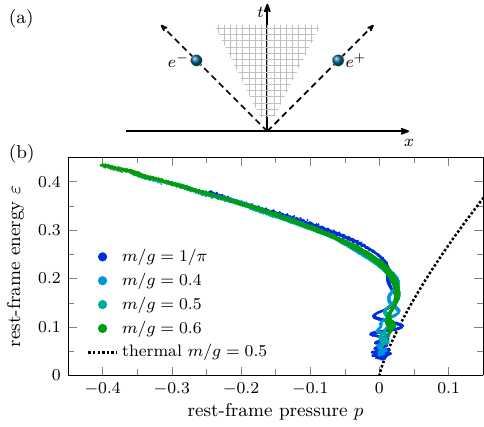}
\end{center}
\caption{Forward light cone in $e^+e^-$ annihilation with a dashed forward wedge, in panel (a). In (b), we show the rest frame energy versus pressure in the forward wedge for $\tau>2.5$ and  $m/g>1/\pi$ in the massive Schwinger model (color). The equilibrium equation of state with $m/g=0.5$ is shown for
comparison (dotted-black).}
\label{fig:EPwedge}
\end{figure}

{\bf Introduction.} Jet fragmentation and hadronization are vigorously pursued at current collider facilities both at RHIC and LHC. They will be sought at the future EIC to extract the partonic structure of matter, the gluon helicity in nucleons, and the mechanism behind the production of diffractive dijets~\cite{Abir:2023fpo,AbdulKhalek:2021gbh}.

The canonical setup for jet fragmentation is $e^+e^-$ annihilation into hadrons. In this process, two luminal jets each carrying a quark and antiquark are locally created through a hard virtual photon. As they recede, a string is formed which subsequently fragments into multi-mesons. Field and Feynman initially described the fragmentation process as an out-in statistical cascade process~\cite{Field:1977fa}. A more phenomenologically but empirically successful model of string fragmentation as an in-out cascade process was captured by the LUND model~\cite{Andersson:1997xwk}. The classical string worldsheet undergoes fragmentation by local pair production inspired by the Schwinger pair-creation process, with the fragmentation function following from
probabilistic considerations involving only the creation points in hyperbolic coordinates on the string worldsheet.

A particularly promising arena for the investigation of these problems is the massive Schwinger model (QED in two dimensions).
In two dimensions, Coulomb law confines, a phenomenon shared by QCD in any dimension. Moreover, the string worldsheet being generically two-dimensional, suggests that much of the physics in the longitudinal fragmentation may be common to QED in two dimensions.

This paper is the first full quantum analysis of the energy-momentum content of the string fragmentation process in the massive Schwinger model between receding luminal external sources. We employ a matrix product state (MPS) representation~\cite{Schollwock:2011review,Orus:2014review,Cirac:2021review}, and simulate the real-time evolution using the time-dependent variational principle (TDVP)~\cite{Haegeman:2011tdvp,Haegeman:2011unyfying}, which enables us to correlate the energy-momentum behavior with the quantum information properties of the process through entanglement entropy. This quantum entropy has been argued to be at the origin of the prompt decoherence in scattering processes at collider energies~\cite{Stoffers:2012mn,Peschanski:2012cw,Kovner:2015hga,Kharzeev:2017qzs,Liu:2018gae,Berges:2017hne,Kutak:2023cwg,Liu:2023eve,Hentschinski:2024gaa}, with much in common with the onset of chaos and the approach to thermalization using chaotic maps~\cite{Latora_1999,Latora:1999vk}, and the rate of emission of entropy by black holes~\cite{Bekenstein:2001qi}.

We note that the fragmentation process in two-dimensional massless QED was originally addressed semi-classically by bosonization~\cite{Casher:1974vf,Kharzeev:2012re}.
Their result shows a conformal single-particle multiplicity.
The same multiplicity was obtained using light-cone quantization with additional assumptions~\cite{Fujita:1989vv}.
More recently, more studies have addressed the issues of string breaking~\cite{Hebenstreit:2013baa,Barata:2023jgd}, and more thoroughly aspects of string breaking,  entanglement and thermalization~\cite{Florio:2023dke,Florio:2024aix,Batini:2024zst}.
For completeness, we also note that partonic physics on quantum computers in two dimensions was explored in~\cite{Lamm:2019uyc,Echevarria:2020wct,Kreshchuk:2020aiq,Perez-Salinas:2020nem,Li:2021kcs,Qian:2021jxp,Li:2022lyt,Zache:2023cfj,Grieninger:2024cdl,Grieninger:2024axp,Barata:2023jgd,Banuls:2024oxa,Kang:2025xpz}, including recent studies of energy correlators~\cite{Barata:2024apg,Lee:2024jnt}.

Our key findings follow from the observation that a qualitative difference in dynamics depends on whether $m/g$ is smaller or greater than $1/\pi$~\footnote{This is not a sharp transition, but rather a crossover.}, which is remarkably close to the location of the spectral crossing noted in~\cite{Grieninger:2024cdl} between the strong and weak coupling regimes, which is also the mass location of the quantum critical point~\cite{ArguelloCruz:2024xzi,Fujii:2024reh} for a vacuum angle $\theta=\pi$. Our findings can be summarized as follows.

For $m/g>1/\pi$, we find a \emph{universal} relation between energy and pressure in the central rapidity region, which has a natural interpretation of an ``equation of state'' (see Fig.~\ref{fig:EPwedge}) of an effective emergent fluid.
We also see a clear separation into two phases of evolution -- ``the string screening phase'' and an out-of-equilibrium  ``free streaming phase'' (characterized by essentially vanishing pressure).
Moreover, the evolution of energy and momentum is in line with the evolution of the entanglement entropy (see Fig.~\ref{fig:EMTEE}).

\begin{figure}[t]
    \includegraphics[width=\columnwidth]{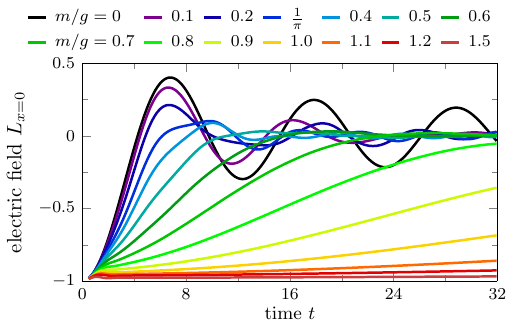}
    \caption{Electric field at mid-rapidity ($x=0$), between two receding luminal jets for a selection of $m/g$.}
    \label{fig:LN}
\end{figure}

For $m/g<1/\pi$, the correlation between energy, momentum, and entanglement entropy rapidly disintegrates, while the effective ``equation of state'' is overshadowed by huge damped oscillations (see, e.g., Fig.~\ref{fig:LN}), characteristic of \emph{classical solutions} of the bosonized formulation of the Schwinger model.

This strongly suggests a crucial role of \emph{quantum} dynamics for $m/g>1/\pi$ and, in particular, for the dynamical appearance of an emergent effective fluid.

\begin{figure}[t]
\includegraphics[width=\columnwidth]{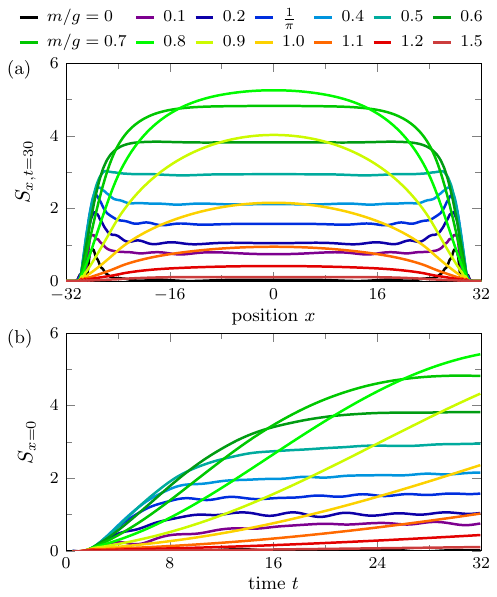}
\caption{
The entanglement entropy between left and right part of the lattice, $S_{x,t}$, where $x$ indicates the position of the cut and $t$ is the time.
In (a), we show the data for $t=30$. In (b), the entanglement entropy between left and right half-spaces with a cut at $x=0$ versus time. In both cases the jet-free entropy is subtracted.}
\label{fig:SEE}
\end{figure}

\begin{figure*}[t]
\includegraphics[width=\linewidth]{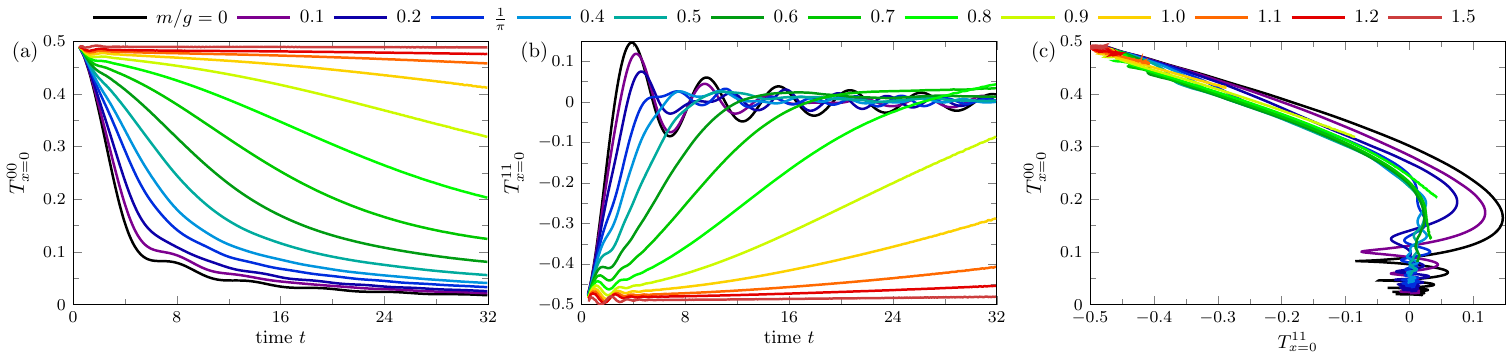}
\caption{The temporal evolution of the energy density $T^{00}$, in (a), and pressure $T^{11}$, in (b), at mid-rapidity $x=0$. In (c), we show the energy versus pressure.}
\label{fig:EMT}
\end{figure*}

{\bf QED2 with jets.}
QED2 with massive fermions is described by~\cite{Schwinger:1962tp,Coleman:1976uz}
\bea
\label{A1}
\mathcal{S} =\int d^2x\,\bigg(\frac 14F^2_{\mu\nu} +\overline \psi (i\slashed{D}-m)\psi\bigg),
\eea
with $\slashed{D}=\slashed{\partial}+ig\slashed{A}$.
The bare fermion mass is $m$ and the coupling $g$ has a mass dimension.
For strong coupling, with $m/g<1/\pi$, the spectrum is composed of three bound and stable mesons~\cite{Coleman:1974bu}.
The weak coupling spectrum, for $m/g>1/\pi$, consists of a tower of bound mesons due to the confining Coulomb law for heavy fermions.

The symmetric energy-momentum tensor (EMT) for the 2D massive Schwinger model is~\cite{Ji:2020bby}
\bea
\label{TMUNUX}
T^{\mu\nu}=\frac 12 E^2g^{\mu\nu}
+\frac i4 \overline\psi\gamma^{[\mu} \overleftrightarrow{D}^{\nu]_+}\psi
\eea
and satisfies the trace identity $T^\mu_\mu=E^2+m\overline\psi\psi$.
In the following, we adopt the standard choice of time-like axial gauge $A^0=0$.

The lattice version of the resulting Hamiltonian follows from the original staggered formulation in~\cite{Banks:1975gq} and is amenable to a tensor network treatment.
After fixing the temporal gauge, one obtains
\eqn
\label{H00}
\mathbb H(t) &=& \f{-i}{2a} \sum_{n=1}^{N-1} \left[ \chid_n \chi_{n+1} -h.c.\right] + \nonumber\\
&& + m \sum_{n=1}^N (-1)^n \chid_n \chi_n +  \f{g^2 a}{2} \sum_{n=1}^N L_n^2
\eqnx
in terms of fermionic variables $\chi_n$ at each site in the staggered formulation, normalized as $\{\chid_n, \chi_m\}=\dl_{nm}$. Here, we consider a finite lattice of $N$ sites with discretized lattice spacing $a$.

The local charge density, including the external jet charges,  fixes the local
electric field $L_n$ by the Gauss law. Here we follow the approach suggested first in~\cite{Florio:2023dke,Florio:2024aix}, where the external charges add
to the dynamical charges through Gauss law constraints
\eq
L_n - L_{n-1} = \chid_n \chi_n - \f{1- (-1)^n}{2} + Q^+_n(t) - Q^-_n(t),
\eqx
which can be solved for $L_n$ in terms of $L_0$ and the charges.
In the tensor network formalism, care must be taken as space is discretized, while time evolution is performed in continuous time. Hence, the trajectories of the charges, $x=\pm v t$, have to be smeared across neighboring sites. A natural way to do this is to set
\bea
Q^\pm_n(t) = Q\cdot \max\left( 1- |n_0 \pm vt/a -n|, 0 \right),
\eea
with $n_0 = (N+1)/2$ corresponding to position $x=0$ at the center of a finite system, assuming $x = a (n - n_0)$ relation between positions and corresponding lattice sites. This ensures that the total charge on each trajectory is $\pm Q$ irrespective of time, and the Hamiltonian does not change abruptly in time despite spatial discretization.
In the present paper, we consider luminal sources, i.e., $v=1$ and unit charges, i.e., $Q=1$.
The time-evolved ground state from the jet-free initial state satisfies the Gauss law in the electric-flux-free state with $L_0=0$.

We perform the temporal evolution for the Schwinger model with masses ranging from the solvable case of $m/g=0$ to $m/g=1.5$, where we fix $g=1$ to set the scale in our simulations.
The results in the main text are obtained for the system of $N=1024$ sites, with discretization $a=1/16$, which translates to positions $|x|<32$ and allows for times $t < 32$ before the jets reach the ends of a finite lattice. To obtain converged results, we run simulations with MPS bond dimensions up to 1024, using MPS that explicitly implements the particle-number conservation of Eq.~\eqref{H00}~(see Supplementary Materials~\cite{SM}).

{\bf EMT and spatial Entanglement Entropy.}
In Fig.~\ref{fig:SEE}(b), we show the entanglement entropy, measured in bits, between the left and right half-spaces with the value for the vacuum subtracted out. In the \emph{massive} case, we see a pattern of increasing entanglement, which then saturates at a plateau level. The linear rise and plateau of the entanglement entropy in time was initially observed for a single mass in~\cite{Florio:2024aix}. It is similar to the rise in information entropy noted in Regge scattering~\cite{Liu:2018gae},
and in general chaotic systems on their way to thermalization~\cite{Latora_1999}, as captured by the Kolmogorov-Sinai entropy rate. In the strong-coupling regime, the onset of chaoticity appears rather early in the jet evolution, and much later at weak coupling.

A similar plateau is seen to develop spatially for the entanglement entropy as a function of the cut position $x$ as a function of $m/g$, between the receding jets, as shown in Fig.~\ref{fig:SEE}(a).
The possible relationship of this plateau with the multiplicities will be addressed elsewhere. We expect a decrease of the plateau region, as we transition from a free boson (strong coupling) to a free fermion (weak coupling) theory asymptotically.

As we show below, those two phases have direct counterparts in the evolution of the EMT. The asymptotic plateau value rises monotonically with the mass parameter of the Schwinger model for the range of masses in Fig.~\ref{fig:SEE}(b). Of particular interest is the massless case, where the entanglement entropy \emph{vanishes} throughout the evolution. This means that the jets and the string breaking-up do not lead to additional entanglement w.r.t. the ground-state vacuum. An important corollary of this observation is that one should not identify the entanglement entropy with particle production in general, as in the massless case the latter is present albeit coherently, see~\cite{Casher:1974vf}, while the former vanishes. The rise of the entanglement originates from the interaction between the produce in the massive case.

\begin{figure*}[t]
\includegraphics[width=\linewidth]{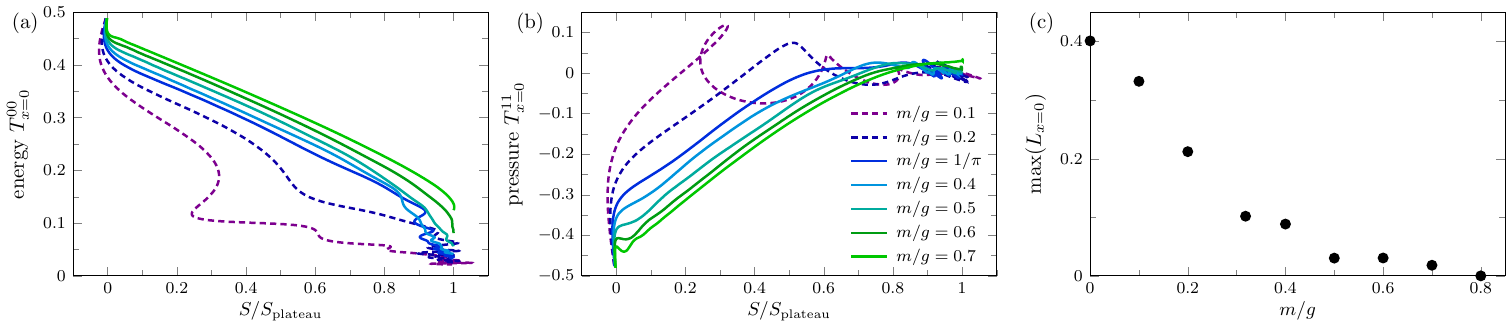}
\caption{The temporal evolution of the energy density $T^{00}$, in (a) and pressure $T^{11}$, in (b), at mid-rapidity $x=0$ as a function of the ratio of entanglement entropy to its plateau value. The dashed lines represent the low $m/g$ case.
In (c), we show the maximal value of the electric field at $x=0$ as a measure of oscillatory behavior.}
\label{fig:EMTEE}
\end{figure*}

The most surprising novel observation of the present paper is a hidden regularity in the temporal evolution of the EMT in the central region of the forward light-cone. In Figs.~\ref{fig:EMT}(a) and (b), we show the temporal evolution of the energy and pressure at mid-rapidity for the Schwinger model with various masses. As by symmetry there is no momentum flow at $x=0$, these coincide with the $T^{00}$ and $T^{11}$ components of the energy-momentum tensor. Despite the wide range of trajectories, energy and pressure are strongly correlated, as shown in Fig.~\ref{fig:EMT}(c).
We observe two distinct qualitative stages of behavior.
The negative pressure, appearing due to the presence of the electric field indicating the stretched string between the receding charges, gets screened and decreases in its magnitude to cross zero, still at a substantially nonzero value of the energy density. This signifies that at this stage matter dominates the screened electric field (compare Fig.~\ref{fig:EMT}(b) with Fig.~\ref{fig:LN}). Then the evolution changes character and we observe a decrease of the energy density at a small pressure (reminiscent of ``free streaming'' or ``dust''). We will elaborate on this interpretation in the following section.

The above qualitative description holds literally for the range of masses $m/g=1/\pi \ldots 0.6$ (for higher masses, the time scale of the evolution is too slow to make a definite statement). For smaller masses, this general picture is swamped by huge oscillations, see Fig.~\ref{fig:EMT}(c).
Indeed, we can quantify the oscillatory character of the dynamics by looking at the maximal \emph{positive} value of the electric field at $x=0$ (compare Fig.~\ref{fig:LN}). This is plotted in Fig.~\ref{fig:EMTEE}(c), where we see a clear decrease of oscillations with increasing mass, with negligible values for higher masses. The mass $m/g \sim 1/\pi$ is roughly in the crossover region, in the vicinity of the spectral crossing from strong to weak coupling~\cite{Grieninger:2024cdl}.

As described above, we also see two distinct qualitative phases in the evolution of the entanglement entropy, as shown in Fig.~\ref{fig:SEE}(a)---a rise followed by saturation at some plateau value. In Figs.~\ref{fig:EMTEE}(a) and (b), we see that for $m/g>1/\pi$, the temporal evolution of energy and pressure is in step with the rise of entanglement entropy. Thus, the increase in entanglement occurs primarily in the ``screening'' phase, while saturation corresponds to the ``free streaming'' one.

For masses lower than $m/g=1/\pi$, we observe a breakdown of the clear link between entanglement entropy and energy and pressure. This is not only due to the huge oscillations, but is more fundamental as can be clearly seen in the massless case where entanglement entropy vanishes, while the evolution of energy and pressure is still significant. Indeed, we know then that the dynamics of the massless Schwinger model is very well described by \emph{classical} evolution of the bosonized form of the theory, with the oscillations following from a Bessel function solution (see Supplementary Materials~\cite{SM}).
This picture is consistent with the lack of entanglement. Conversely, the clear synchronization between the entanglement entropy and energy/pressure for $m/g>1/\pi$, as well as the lack of oscillations, very strongly suggests that \emph{quantum} dynamics is crucial for the string breaking physics of the Schwinger model in this range of masses.

{\bf Emergent perfect fluid.} The simple relation between energy and pressure at mid-rapidity, shown in Fig.~\ref{fig:EMT}(c) for $m/g>1/\pi$, is reminiscent of an ``equation of state'' for a fluid. In this section, we will explore further this interpretation.
A general form of a hydrodynamic tensor is
\eq
\label{e.hydro}
T^{\mu\nu} = (\eps + p) u^\mu u^\nu - p \eta^{\mu\nu} + \Pi^{\mu\nu},
\eqx
where $\Pi^{\mu\nu}$ contains derivatives of the flow velocity and can be chosen to satisfy $\Pi^{\mu\nu} u_{\nu}=0$. In two dimensions, the number of independent components of EMT is exactly equal to the number of hydrodynamic variables: energy, pressure, and flow velocity. Hence, \emph{any} two-dimensional energy-momentum tensor can be always \emph{formally} expressed in a perfect-fluid hydrodynamic form (i.e., with $\Pi^{\mu\nu}=0$).
Therefore, to make a case for the emergence of a fluid behavior, we also have to consider some additional physical arguments.

\begin{figure}[t]
    \centering
    \includegraphics[width=\columnwidth]{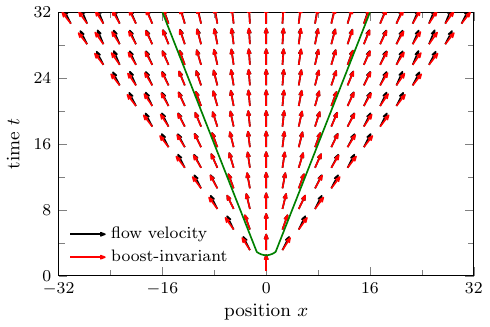}
    \caption{Hydrodynamic flow velocity in the forward light cone (black arrows) for $m/g=0.6$. The boost-invariant flow is marked in red.  The green line delineates the forward wedge with $|x| < 0.5 t$ and $\tau > 2.5$.}
    \label{fig:velocity}
\end{figure}

In particular, for the hydrodynamic interpretation to make sense, the resulting energy $\eps$ and pressure $p$ should be linked by a physically sensible equation of state.
The (rest frame) energy density $\eps$ and fluid velocity $u^\mu$, can be found by solving $T^{\mu\nu}u_\nu = \eps u^\mu$. In Fig.~\ref{fig:velocity}, we show the velocity compared to the boost-invariant flow. We also mark, somewhat arbitrarily, a forward region defined by $|x|< 0.5\, t$, restricting also to $\tau>2.5$ as we do not expect to have a fluid interpretation at very early times and close to the receding jets.

For now, let us neglect the dissipative part $\Pi^{\mu\nu}$. Then we can make a scatter plot of $\eps$ versus $p$ in the entire forward region. The result is shown in Fig.~\ref{fig:EPwedge}(b). We observe a universal relation consistent with the interpretation of an effective ``equation of state''. Note that even in the ``free streaming phase'', the $\eps$-$p$ relation departs from the equilibrium equation of state in the strong coupling phase with $m/g=0.5$, as shown in Fig.~\ref{fig:EPwedge}(b) with the dotted black line. The produced phase with nearly null pressure is thus out-of-equilibrium.

Now, suppose that we turn on dissipation. Then $p$ shown in Fig.~\ref{fig:EPwedge}, would really be $p+\Pi^{00}$. But, as $\Pi^{\mu\nu}$ incorporates velocity gradients, which for boost-invariant flow would have explicit $1/\tau$ dependence, this would likely upset the nice single curve seen in Fig.~\ref{fig:EPwedge} as the points $(\eps, p)$ for various masses occur with different $\tau$. Therefore, we expect that possible dissipative terms are rather small, at least in the forward region that we consider, indicative of an approximately ``perfect'' fluid.

{\bf Discussion.}
Using tensor network methods, we have analyzed $e^+e^-$ jet fragmentation in massive two-dimensional QED with a focus on the evolution of the energy-momentum tensor and the spatial entanglement entropy. For $m/g>1/\pi$ close to the spectral crossing from strong to weak coupling, an effective fluid behavior sets in around the mid-rapidity region, that is strongly correlated with the rise of the entanglement entropy, with a universal energy-pressure curve.

However, this emergent fluid is quite nonstandard, as we do not expect the energy and pressure appearing in the equation of state to be parametrized by a standard temperature. Also, this effective fluid is not charged, as the presence or absence of an electric field does not influence the flow velocity at mid-rapidity, where the fluid is at rest. So, it is rather a ``dipole'' or ``mesonic'' fluid, which appears to be out of equilibrium. An analysis of the multiplicities in this region will be discussed elsewhere. Remarkably, a hydrodynamical description of high-energy $e^+e^-$ annihilation into hadrons was explored awhile ago using Landau's arguments~\cite{Shuryak:1971hn}.

The correlation between the spatial entanglement entropy and the evolution of the energy and pressure for $m/g>1/\pi$, when compared to the lack thereof and the huge oscillations reminiscent of classical dynamics for $m/g<1/\pi$, is a clear indication of \emph{quantum} dynamics for the jet-fragmentation process for $m/g>1/\pi$, near the separatrix between strong and weak couplings. The emergent universal energy-pressure relation, noted in this work, calls for a better understanding of the underlying physics at work in this transition region.

Finally, this emergent chargeless and nearly perfect fluid between luminally receding jets at mid-rapidity produced by string breaking, might be described holographically by a ``falling wormhole'' in bulk using the principle of maximum string entropy~\cite{Susskind:2005js}, if the sourcing jets are identified with the holographic Schwinger pair in~\cite{Jensen:2013ora,Sonner:2013mba,Grieninger:2023ehb,Grieninger:2023pyb} (although in the latter the string breaking process is absent).
Also, a similar observation for a fireball formation at mid-rapidity was noted in high-energy scattering in the Regge limit using holography~\cite{Basar:2012jb,Stoffers:2012mn,Qian:2015boa,Shuryak:2017phz,Liu:2018gae}.
In the latter, an emergent temperature was identified with the ``Unruh temperature'' induced by the rapidity change on the exchanged string hyperboloid worldsheet, with a small primordial viscosity~\cite{Qian:2012vuf}.
This mechanism may be at work in the present case for high-multiplicity fragmentation events. This emergent fluid may be analyzed using energy-energy or energy-charge correlators, or through select multi-particle correlations and their moments at mid-rapidity, at current collider facilities.

\acknowledgments{
{\it Acknowledgements.} IZ thanks the Marc Kac center in Krakow, RJ and MAN thank Stony Brook University for hospitality during the initial stages of this work.
IZ  work is supported in part by the Office of Science, U.S. Department of Energy under Contract No. DE-FG-88ER40388.
RJ, MAN and MMR are supported in part by a  Priority Research Area DigiWorld grant under the Strategic Programme Excellence Initiative at the Jagiellonian University (Kraków, Poland).
MMR acknowledges support by the National Science Center (NCN), Poland, under projects no. 2020/38/E/ST3/00150.
}

%

\pagebreak

\appendix
\onecolumngrid
\setcounter{equation}{0}
\setcounter{figure}{0}
\renewcommand{\theequation}{S\arabic{equation}}
\renewcommand{\thefigure}{S\arabic{figure}}

\begin{center}
\textbf{\large Supplemental Materials}
\end{center}

\section{Energy-momentum tensor}

Using the staggered fermion formulation~\cite{Kogut:1974ag} and the trace identity, we can rewrite the components of the energy-momentum tensor (EMT) in Eq.~(2) of the main text as follows
\eqn
\label{TMUNUBIT}
T^{00} &=& \f{-i}{2a^2} \left[ \chid_n \chi_{n+1} -hc\right] +\f{m}{a} (-1)^n \chid_n \chi_n + \f{g^2}{2} L_n^2, \nonumber \\
T^{11} &=& \f{-i}{2a^2} \left[ \chid_n \chi_{n+1} -hc\right] - \f{g^2}{2} L_n^2, \nonumber\\
T^{01} &=& \f{-i}{4a^2} \left[ \chid_n \chi_{n+2} - \chid_{n+2} \chi_{n}\right].
\eqnx
The $T^{00}$ component in~\eqref{TMUNUBIT}, which is the time-dependent Hamiltonian in Eq.~(3) of the main text,
can be used  to evolve from the initial  jet-free ground state~\cite{Florio:2023dke},
\bea
\label{PSIT}
|\Psi(t)\rangle=
{\cal T}{\rm exp}\bigg(-i\int_0^t\,dt'\,\mathbb H(t')\bigg) |\Psi(t=0)\rangle,
\eea
with Gauss law resolved. Note that the numerical results for the EMT induced by the luminal jets in the main text have the jet-free ground-state contribution removed.

\section{Hydrodynamical variables}

In two dimensions, the number of independent components of the energy-momentum tensor is exactly equal to the number of hydrodynamic variables: energy, pressure, and flow velocity. Hence any two-dimensional energy-momentum tensor can be always formally expressed in a perfect-fluid hydrodynamic form,
\eq
\label{e.perfectfluid}
T^{\mu\nu} = (\eps + p) u^\mu u^\nu - p \eta^{\mu\nu},
\eqx
after dropping potential dissipative corrections $\Pi^{\mu\nu}$ in Eq.~(6), which are small as we argued in the main text. With this in mind, the hydrodynamical variables in~\eqref{e.perfectfluid} follow as
\eqn
\eps &=& \f{1}{2} \left( T^{00} - T^{11} + \sqrt{(T^{00} +T^{11})^2 - 4 (T^{01})^2} \right),\nonumber \\
p &=& \eps - T^{00} +T^{11}, \nonumber\\
u^\mu &\propto& \left(1, \f{2 T^{01}}{T^{00} + T^{11} + \sqrt{(T^{00} +T^{11})^2 - 4 (T^{01})^2} } \right),
\nonumber\\
\eqnx
with the flow velocity $u^1$ at large times shown in Fig.~6 of the main text.

\section{Massless QED2 analytical analysis}

The massless results presented in the main text have been explicitly checked by analytical methods. Indeed,
massless QED2 in the strong coupling maps onto a massive free boson.  Using standard 2D-bosonization
$j_\mu=\epsilon_{\mu\nu}\partial^\nu\varphi/\sqrt\pi$~\cite{Coleman:1974bu}, the EMT in Eq.~(2) reads
\bea
\label{TBOSON}
T^{\mu\nu}=\frac 12 \partial^{[\mu}\varphi\,\partial^{\nu]_+}\varphi
-\frac 12 g^{\mu\nu}(\partial_\lambda\varphi\partial^\lambda\varphi-m_S^2\varphi^2).
\eea
When jet sources are added, they produce a constant electric field
in the forward part of the light cone, fixed by the external field
$\varphi_{ext}=-\sqrt{\pi} \th(\tau^2)$ with squared proper time $\tau^2=t^2-x^2$. The classical solution induced by
this time-varying external source was worked out in~\cite{Casher:1974vf,Kharzeev:2012re},
\eq
\label{CLASS}
\varphi(\tau) = \sqrt{\pi} \left( 1- J_0(m_S\tau) \right) \th(\tau^2),
\eqx
with the vector current components
\eqn
j^0(t,x) &=& -m_S \f{x}{\tau} J_1(m_S \tau)\th(\tau^2), \\
j^1(t,x) &=& -m_S \f{t}{\tau} J_1(m_S \tau)\th(\tau^2).
\eqnx
The corresponding components of the EMT are
\eqn
\label{EMTZ}
T^{00}(t,x) &=& \sigma_T \left( J_0^2(m_S \tau)  + \f{t^2+x^2}{\tau^2} J_1^2(m_S \tau)  \right)\th(\tau^2), \nonumber\\
T^{11}(t,x) &=& \sigma_T \left( -J_0^2(m_S \tau)  + \f{t^2+x^2}{\tau^2} J_1^2 (m_S \tau) \right) \th(\tau^2), \nonumber\\
T^{01}(t,x) &=& \sigma_T \f{2tx}{\tau^2} J_1^2(m_S \tau) \th(\tau^2),
\eqnx
with the string tension $\sigma_T=\pi m_S^2/2$, and  satisfy
\bea
\label{EMTFLOW}
T^{\mu\nu}u_\nu = \varepsilon u^\mu,
\eea
with the time-like 4-velocity $u^\mu=x^\mu/\tau$ and
\bea
\label{EFLOW}
\varepsilon(\tau)=\sigma_T (J_0^2(m_S\tau) +J_1^2(m_S\tau))\th(\tau^2).
\eea
The latter is the energy density in the co-moving frame at mid-rapidity. The electric field between the receding jets is
\bea
\label{LELEC}
L(\tau)=-m_S(\varphi+\varphi_{ext})= \sqrt{2\sigma_T}\,J_0(m_S \tau)\,\theta(\tau^2).
\eea
As noted earlier, when comparing~\eqref{EMTZ} to the numerical results for the massless case in Fig.~\ref{fig:S2} below, a subtraction of the jet-free ground state contribution is carried out.

\begin{figure*}[t]
\includegraphics[width=\linewidth]{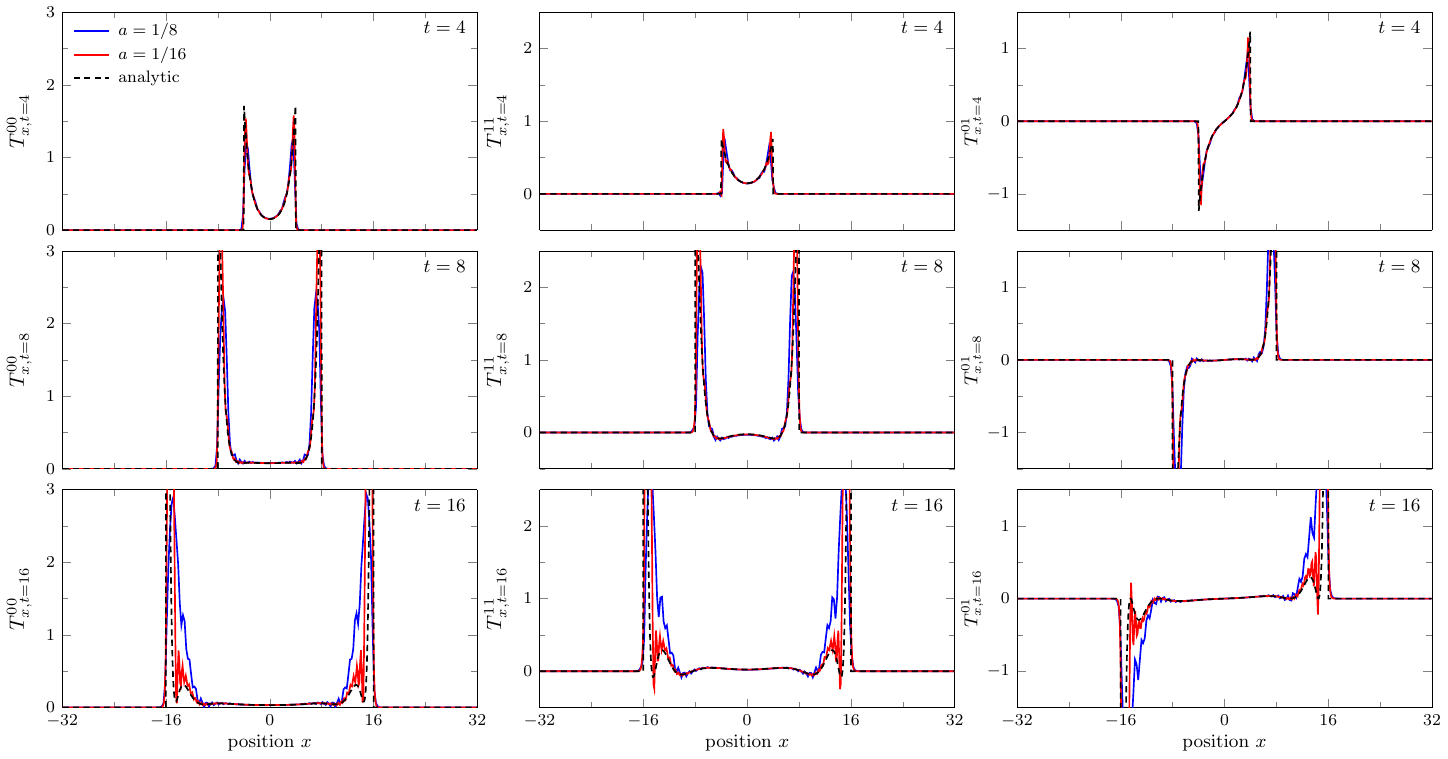}
\caption{Comparison of the components of the energy momentum tensor versus space for different times $t=4,8,16$. The analytic results for $m/g=0$ (dashed-black) compare well with the numerical results for lattice spacing $a=1/8$ with $N=512$ (solid-blue) and $a=1/16$ with $N=1024$ (solid-red) in the mid-rapidity region.}
\label{fig:S2}
\end{figure*}

\begin{figure*}[t]
\includegraphics[width=\linewidth]{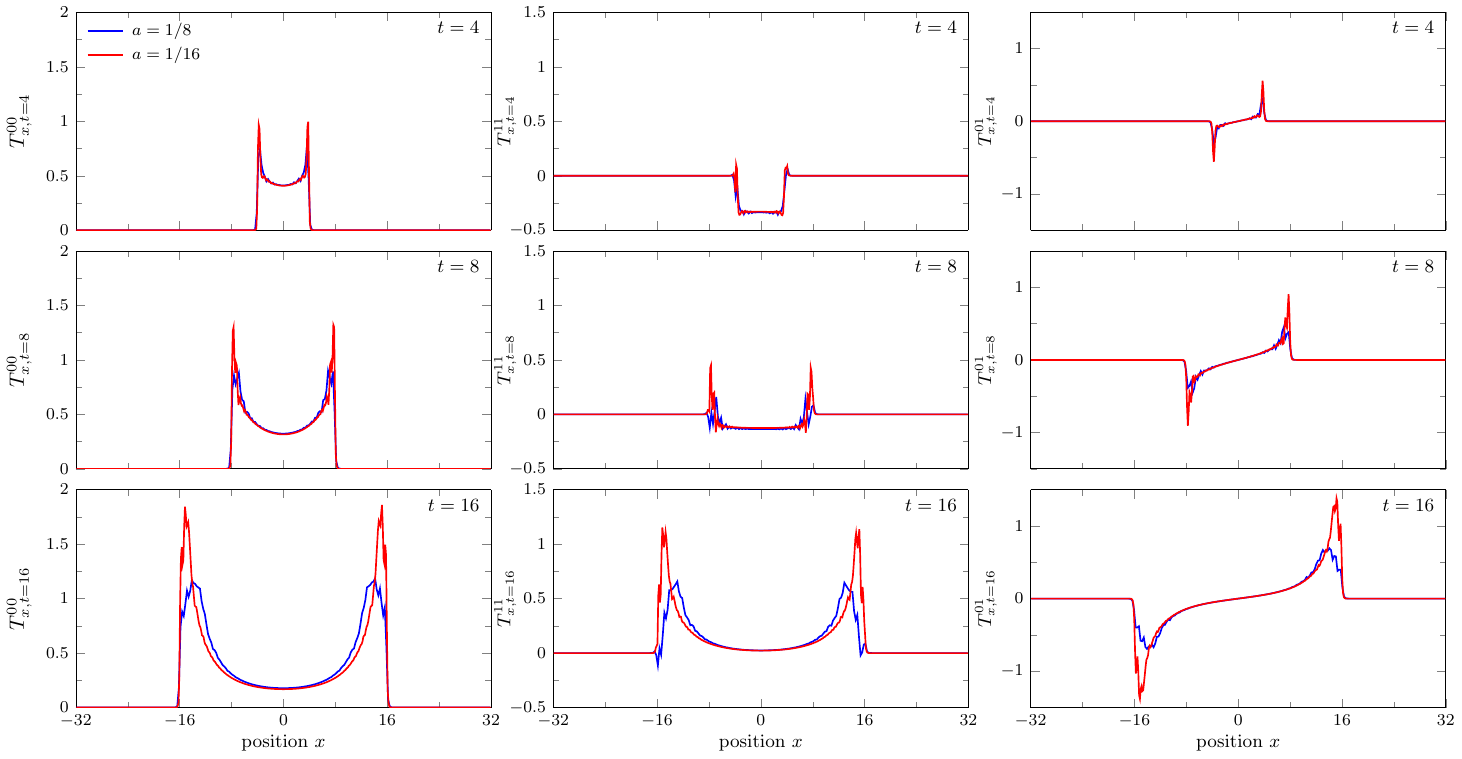}
\caption{Comparison of the components of the energy momentum tensor versus space for different times $t=4,8,16$. The  numerical results for $m/g=0.6$ with lattice spacing $a=1/8$ with $N=512$ (solid-blue) and $a=1/16$ with $N=1024$ (solid-red), show good convergence in the mid-rapidity region.}
\label{fig:S3}
\end{figure*}

QED2 in the massless case is also amenable to the Sugawara form, a re-casting solely in terms of the currents~\cite{Sugawara:1967rw},
\bea
\label{TMUNUS}
T^{\mu\nu}=\frac {E^2}2\,g^{\mu\nu}
+\frac \pi 2(j^\mu j^\nu+j^\nu j^\mu -g^{\mu\nu} j_\lambda j^\lambda),
\eea
with $E$ fixed by Gauss law, and modulo normal ordering. In terms of the classical solution~\eqref{CLASS}, the charge currents are
\bea
j^\mu(t,x)=\frac{\varphi'(\tau)}{\sqrt{\pi}}\,\tilde u^\mu
 \equiv \rho(\tau) \tilde u^\mu,
\eea
with dual velocity $\tilde u^\mu=\epsilon^{\mu\nu}u_\nu$. The charge currents flow in dual space in contrast to the energy flow in space in~\eqref{EMTFLOW}, with $\partial_\mu (g\rho u^\mu+gj^\mu_Q)=0$
by charge conservation (the axial charge is non-conserved).
\eqref{TMUNUS}~maps onto the EMT of an {\it apparent ideal fluid} with dual velocities
\bea
\label{TMUNUF}
T^{\mu\nu}= \pi \rho^2\,\tilde u^\mu \tilde u^\nu+\frac 12 (E^2+\pi \rho^2)\,g^{\mu\nu}.
\eea

In Fig.~\ref{fig:S2} we compare the massless analytic results for the energy momentum tentor components (black-dashed curve) with the jet-free part subtracted, to the numerical results obtained by the tensor network method versus the space separation for different times.
The numerical results at the lattice spacing $a=1/8$ (blue solid curve) and $a=1/16$ (red solid curve) show good agreement with the analytical results in the mid-rapidity region.
The signals at the jets are growing rapidly with time, requiring finer discretization. However, in the main text we focus on the mid-rapidity region with $|x| < 0.5 t$, where the results are well converged.
The same observation holds for the massive case. In Fig.~\ref{fig:S3}, we illustrate the comparison for the same components of the EMT for $m/g=0.6$ for which there is no analytical result. Again, we observe good convergence of the results with smaller lattice spacing.

\section{Entanglement entropy, convergence checks and further simulation details}

For a pure state $|\Psi(t)\rangle$ supported on $N$ sites, the reduced density matrix of that state supported on the first $n$ modes is
\begin{equation}
    \rho(t)_{[1,\ldots, n]} = \rm{Tr}_{[n+1, n+2, \ldots, N]} |\Psi(t) \rangle \langle \Psi(t)|,
\end{equation}
where the partial trace is performed over the last $N-n$ modes.
The bipartite entropy follows as
\bea
S_{x,t} = -\rm{Tr}\left(\rho(t)_{[1,\ldots, n]} \log \rho(t)_{[1,\ldots, n]}\right),
\eea
where we use base-$2$ logarithm in our simulations
and identify the position of the cut $x$ through $x = a (n-n_0)$, with $n_0=(N+1)/2$ and the lattice spacing  $a$. The bipartite entanglement entropy is readily accessible from the MPS representation~\cite{Schollwock:2011review, Orus:2014review, Cirac:2021review}. Indeed, in a pure state it directly follows from a Schmidt decomposition of $|\Psi(t)\rangle$,
\begin{equation}
|\Psi(t)\rangle = \sum_i \lambda_i |\Psi_{[1,\ldots, n]}^i \rangle|\Psi_{[n+1,\ldots, N]}^i \rangle,
\end{equation}
where $\lambda_i$ are Schmidt values related to the $n$-th cut, and $|\Psi_{[1,\ldots, n]}^i \rangle$ ($|\Psi_{[n+1,\ldots, N]}^i\rangle $) are the corresponding Schmidt vectors supported on the first $n$ (the last $N-n$) modes. The MPS approximation is based on the assumption that one can faithfully represent the state by retaining $D$ dominant Schmidt values, where $D$ is the MPS bond dimension~\cite{Schollwock:2011review, Orus:2014review, Cirac:2021review}. The von Neuman entropy follows as
\begin{equation}
    S_{x,t} = -\sum_{i=1}^D \lambda_i^2 \log \lambda_i^2.
\end{equation}

\begin{figure}[t]
\includegraphics[width=0.5 \linewidth]{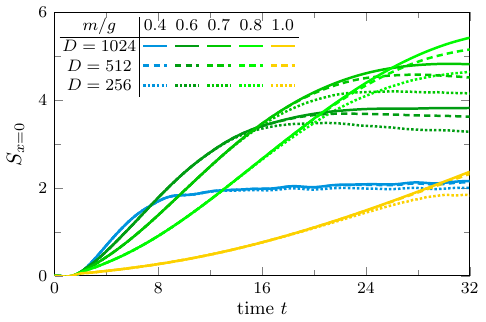}
\caption{Entanglement entropy at mid-rapidity as a function of time,
for different masses $m/g$ and bond dimensions $D$. The jet-free entropy is subtracted.}
\label{fig:S1}
\end{figure}

The entanglement entropy provides a sensitive measure of convergence. In Fig.~\ref{fig:S1}, we show the results at mid-rapidity $x=0$, with the ground state contribution subtracted, as a function of time for different masses $m/g$. We focus on the vicinity of $m/g=0.7$, which is the most challenging regime for our parameter range. With the bond dimensions reaching $D=1024$ we obtain a good convergence.

Finally, we comment on some aspects of our numerical simulations, for which we employ the YASTN tensor network library~\cite{yastn}. YASTN can work directly with fermionic variables, so an explicit Jordan-Wigner transformation of the EMT operators in Eq.~\eqref{TMUNUBIT} and the Hamiltonian in Eq.~(3) of the main text to spin variables is not necessary. The initial ground state is obtained with a standard density matrix renormalization group (DMRG) procedure~\cite{Schollwock:2011review}. We explicitly employ the conservation of the number of $\chi$-fermions in our simulations, where the ground state belongs to the space with $N/2$ particles (we consider even $N$ only). For a subsequent time evolution, we employ the time-dependent variational principle (TDVP)~\cite{Haegeman:2011tdvp,Haegeman:2011unyfying}. We truncate the Schmidt values to a relative tolerance $10^{-6}$, or up to a maximal bond dimension $D$, whichever is more restrictive. For an inherently inhomogeneous process, we consider, where large bond dimensions are necessary only for bonds in between the two receding jets, we combine the standard TDVP 1-site procedure (which is faster) and the 2-site procedure (which allows increasing bond dimensions)~\cite{Haegeman:2011tdvp,Haegeman:2011unyfying}. Namely, we follow a strategy in which 2-site updates are executed only {\it locally} during each time step sweep if growth of a given bond dimension is possible based on the tolerance and the maximal bond dimension (see the Supplementary Material of~\cite{rams:2020breaking} for details). A time step $dt=1/16$ with a second-order TDVP method is sufficient for convergence. The finite-temperature simulation in the Fig.~1 of the main text is obtained by imaginaty-time simulation the of density-matrix purification.
Finally, to minimize the finite discretization effects of the staggered fermion representation, we average local EMT terms between even and odd $n$.

\end{document}